\documentclass[fleqn,usenatbib]{mnras}

\usepackage[T1]{fontenc}
\usepackage{ae,aecompl}

\usepackage{graphicx}	% Including figure files
\usepackage{amsmath}	% Advanced maths commands
\usepackage{amssymb}	% Extra maths symbols

\title[A halo L3 subdwarf in the Galactic plane]{Primeval very low-mass stars and brown dwarfs -- V. \\ A halo L3 subdwarf with prograde eccentric orbit in the Galactic plane}

\author[Z. H. Zhang et al.]{Z. H. Zhang,$^{1}$\thanks{E-mail:
zenghuazhang@hotmail.com}\thanks{PSL fellow} A. J. Burgasser$^{2}$ and L. C. Smith$^{3,4}$\\
$^{1}$GEPI, Observatoire de Paris,  Universit{\'e} PSL, CNRS, 5 Place Jules Janssen, F-92190 Meudon, France \\
$^{2}$Center for Astrophysics and Space Science, University of California San Diego, La Jolla, CA 92093, USA \\
$^{3}$Institute of Astronomy, University of Cambridge, Madingley Road, Cambridge CB3 0HA, UK \\
$^{4}$School of Physics, Astronomy and Mathematics, University of Hertfordshire, College Lane, Hatfield AL10 9AB, UK
}

\date{Accepted 2019 March 5. Received 2019 February 26; in original form 2018 October 8}
\pubyear{2019}

\begin{document}
\label{firstpage}
\pagerange{\pageref{firstpage}--\pageref{lastpage}}
\maketitle

\begin{abstract}
   VVV J12564163$-$6202039 is an L subdwarf located in the Galactic plane ($b$ = 0\fdg831) discovered by its high proper motion (1.1 arcsec yr$^{-1}$).  We obtained an optical to near-infrared spectrum of it with the X-shooter on the Very Large Telescope, and re-classified it as an sdL3 subdwarf. Its best-fitting BT-Dusty model spectrum has $T_{\rm eff}$ = 2220 K, [Fe/H] = $-0.9$, and log $g$ = 5.5. It is just below the stellar/substellar boundary in the $T_{\rm eff}$ versus [Fe/H] parameter space. This object is at a distance of 66.94$^{+8.49}_{-6.77}$ pc according to the {\sl Gaia} astrometry. It has a halo membership probability of 99.82 per cent according to its Galactic kinematics. It has unusually flat (|$Z$| $\la$ 500 pc) and extremely eccentric (e = 0.9) prograde orbit that takes the source from as close as 1.2 kpc from the Galactic centre to 24 kpc out. VVV J12564163$-$6202039 joined by SDSS J133348.24+273508.8 and ULAS J021258.08+064115.9 are the only three known objects in the sdL subclass that have halo kinematics. 
\end{abstract}


%%%%%%%%%%%%%%%%%%%%%%%%%%%%%%%%%%%%%%%%%%%%%%%%%%

%%%%%%%%%%%%%%%%% BODY OF PAPER %%%%%%%%%%%%%%%%%%

\section{Introduction}
L subdwarfs are composed of metal-deficient very low-mass stars (VLMS) and transitional brown dwarfs (BD) of the Galactic halo and thick disc. Generally, early-type L subdwarfs are VLMS and mid-type L to early-type T subdwarfs are all transitional BDs (T-BD). T-BDs have unsteady hydrogen fusion in their cores, and the efficiency/intensity of this fusion is sensitive to their masses \citep{zha18b}. The BD transition zone covers a small mass range but spans wide ranges of luminosity, temperature ($T_{\rm eff}$), and spectral type for old thick disc or halo populations \citep{zha18b}. This is one of the reasons that L subdwarfs are so rare, apart from low fractions of thick disc and halo populations in the solar neighbourhood. 

The first known L subdwarf, 2MASS J05325346+8246465 (2M0532) was initially selected as a T dwarf candidate for its blue near infrared (NIR) colour but spectroscopically confirmed as a metal-deficient L subdwarf \citep{bur03}.
There are about 66 L subdwarfs known in the literature to date \citep[see table 1 of][]{zha18b}. One-third of them are substellar objects \citep[e.g. 2M0532;][]{burg08} and two-thirds are VLMS. Since L subdwarfs are faint and emit most of their flux in the NIR, only 20 of known L subdwarfs are detected in the second data release (DR2) of the European Space Agency's {\sl Gaia} astrometric satellite \citep{gaia18}.  

Known L subdwarfs are usually identified in wide field surveys off the Galactic plane to avoid contamination of the disc population. However, an L subdwarf, VVV J12564163$-$6202039 (VVV 1256$-$6202) was recently discovered in the Galactic plane ($l$ = 303\fdg547; $b$ = 0\fdg831) by its high proper motion. It has a proper motion of $\mu_{\rm RA}$ =  $-$1116.3$\pm$4.1 mas yr$^{-1}$  and $\mu_{\rm Dec}$ =  3.9$\pm$4.0 mas yr$^{-1}$ according to the VVV Infrared Astrometric Catalogue \citep[VIRAC;][]{smit18}. The VIRAC is based on observations of the Visible and Infrared Survey Telescope for Astronomy's \citep[VISTA;][]{suth15} Variables in the Via Lactea survey \citep[VVV;][]{minn10}.

VVV 1256$-$6202 was classified as an sdL7 subdwarf based on a poor-quality NIR spectrum \citep[fig. 17 of][]{smit18}. Meanwhile, its optical to NIR photometric colours ($i-J$ and $J-K$) indicate an earlier spectral subtype. It was difficult to tell what caused the inconsistency of its spectral type inferred from its NIR spectrum and photometric colours before we get a good quality spectrum of it. 

VVV 1256$-$6202 is the faintest known L subdwarf that is detected in the {\sl Gaia} DR2. It is detected in the {\sl Gaia} $G$ band (20.75 mag) but not in the $G_{\rm RP}$ or $G_{\rm BP}$ band. The {\sl Gaia} DR2 measured its parallax to be 14.94$\pm$1.68 corresponding to a distance of 66.94$^{+8.49}_{-6.77}$ pc. It has a proper motion of $\mu_{\rm RA}$ =  $-$1129.81$\pm$3.25 mas yr$^{-1}$  and $\mu_{\rm Dec}$ =  21.29$\pm$2.56  mas yr$^{-1}$ in {\sl Gaia} DR2. It has a tangential velocity of 359$^{+45}_{-36}$ km s$^{-1}$ which suggests a halo membership \citep[see fig. 23 of][]{zha18b}. It also has a small $W$ Galactic velocity (perpendicular to the Galactic plane) despite its unknown radial velocity (RV).  

VVV 1256$-$6202 is certainly an unusual and very interesting object that is worth further investigation. A good-quality medium-resolution optical to NIR spectrum of VVV 1256$-$6202 is required to classify its spectral type, and measure its RV and atmospheric properties. Then we could check whether it is a T-BD by its $T_{\rm eff}$ and metallicity ([Fe/H]), and calculate its kinematics and Galactic orbit by its RV and {\sl Gaia} astrometry. 

This is the fifth paper of a series titled {\sl Primeval very low-mass stars and brown dwarfs}. The first four papers of the series are focused on the discovery and characterization of L subdwarfs, as well as discussions on the stellar/substellar boundary, T-BDs, and the BD transition zone \citep{zha17a,zha17b,zha18a,zha18b}. Population properties of metal-poor degenerate brown dwarfs are discussed in the sixth paper \citep{zha19}. Some of these works are summarized in \citet{zha18c}. In this paper we present an optical to NIR spectrum, characterization, and Galactic orbit of VVV 1256$-$6202. Observations are presented in Section \ref{sobs}. Spectral classification, atmospheric properties, and Galactic orbit are presented in Section \ref{scha}. Our conclusion is presented in Section \ref{scon}.

\section{VLT X-shooter spectroscopy}
\label{sobs}

We obtained an optical to NIR spectrum of VVV 1256$-$6202 using the X-shooter spectrograph \citep{ver11} on the Very Large Telescope (VLT) on 2018 May 25 under a seeing of 0.97 and an airmass of 1.28. The X-shooter spectrum was observed in an ABBA nodding mode with a 1.2 arcsec slit providing a resolving power of 6700 in the VIS arm and 4000 in the NIR arm. We break up our total exposure into 10 single integrations of 285 s for the VIS arm and 300 s for the NIR arm. The spectrum was reduced to a flux-calibrated 2D spectrum with the ESO Reflex \citep{freu13}. Then we extracted the spectrum to a 1D spectrum with the {\scriptsize IRAF}\footnote{{\scriptsize IRAF} is distributed by the National Optical Observatory, which is operated by the Association of Universities for Research in Astronomy, Inc., under contract with the National Science Foundation.} task {\scriptsize APSUM}. Telluric correction was achieved using a B9V telluric standard (HIP 79237) which was observed three hours after our target at an airmass of 1.05. 

\begin{figure}
\begin{center}
   \includegraphics[width=\columnwidth]{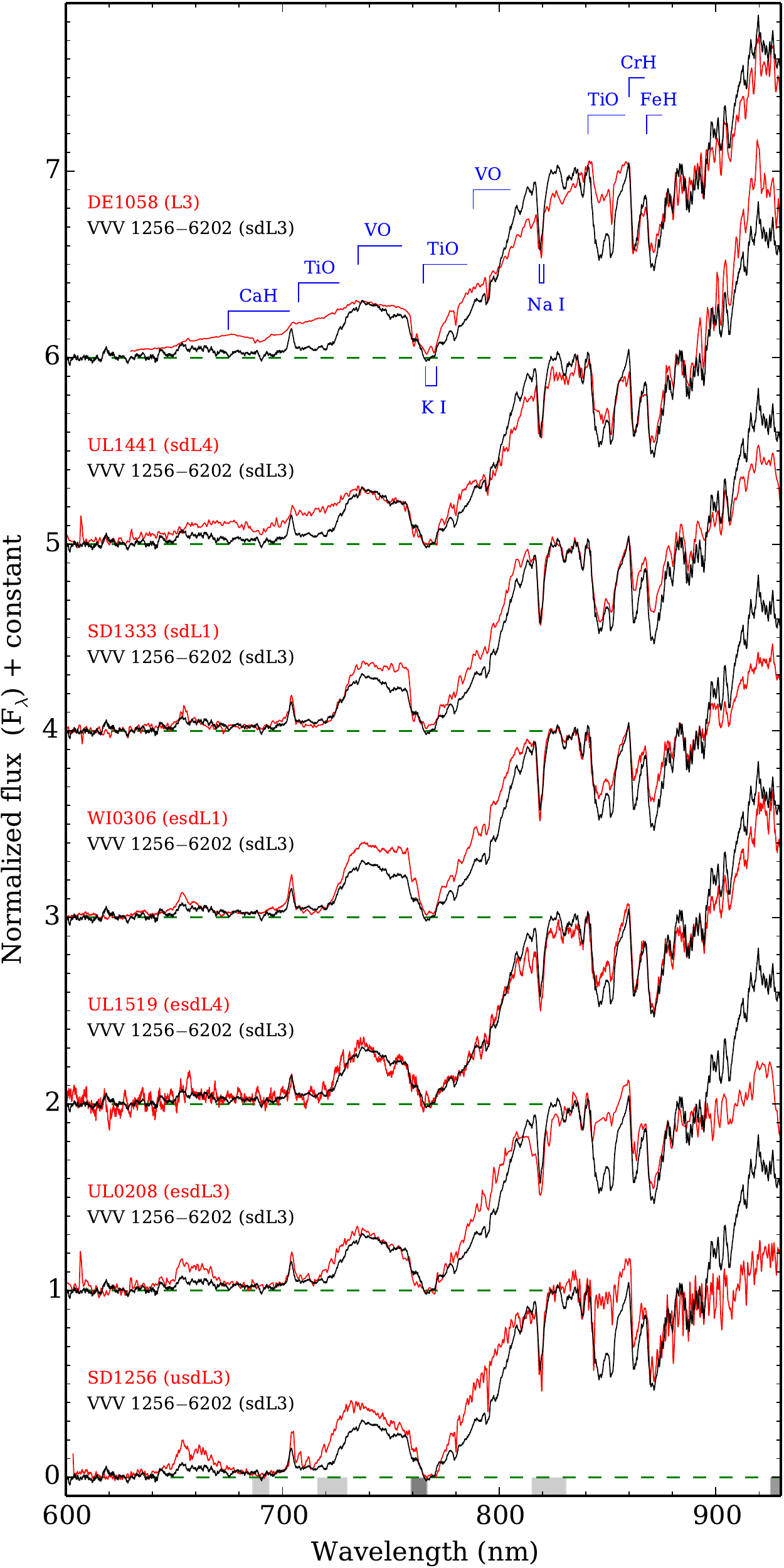}
\caption[]{Optical spectrum of VVV 1256$-$6202 compared to that of DE1058 \citep{kir99}, UL1441 \citep{zha18b}, SD1333, UL1519 \citep{zha17a}, UL0208 \citep{zha18a}, SD1256 \citep{bur09}, and WI0306. Spectra are normalized at 840 nm. }
\label{fvis}
\end{center}
\end{figure}

\begin{figure}
\begin{center}
   \includegraphics[width=\columnwidth]{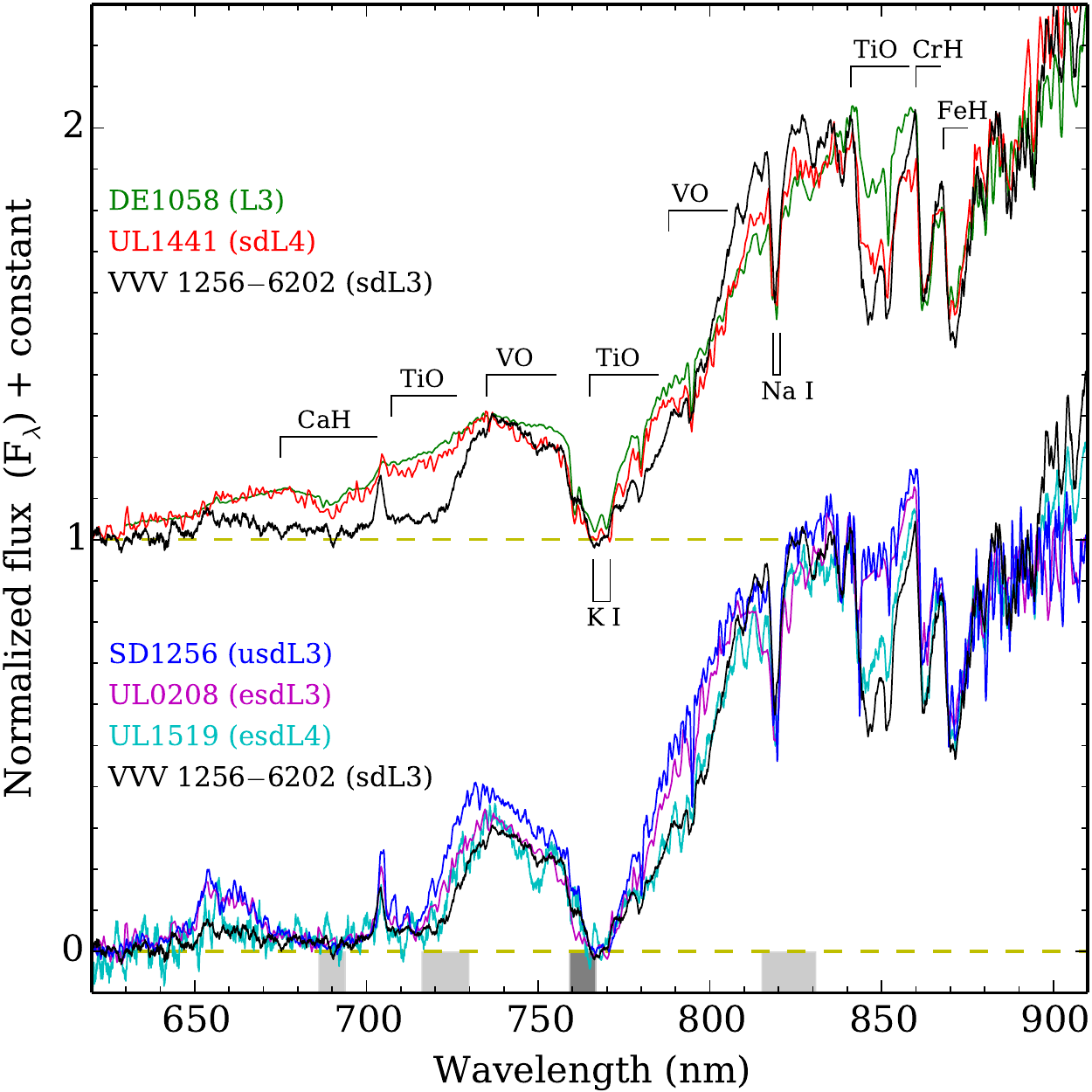}
\caption[]{Strength change of the TiO absorption band at 850 nm across L3, sdL3, esdL3, and usdL3 subclasses with reducing metallicity. Spectra are normalized at 840 nm. }
\label{ftio}
\end{center}
\end{figure}

\begin{figure}
\begin{center}
   \includegraphics[width=\columnwidth]{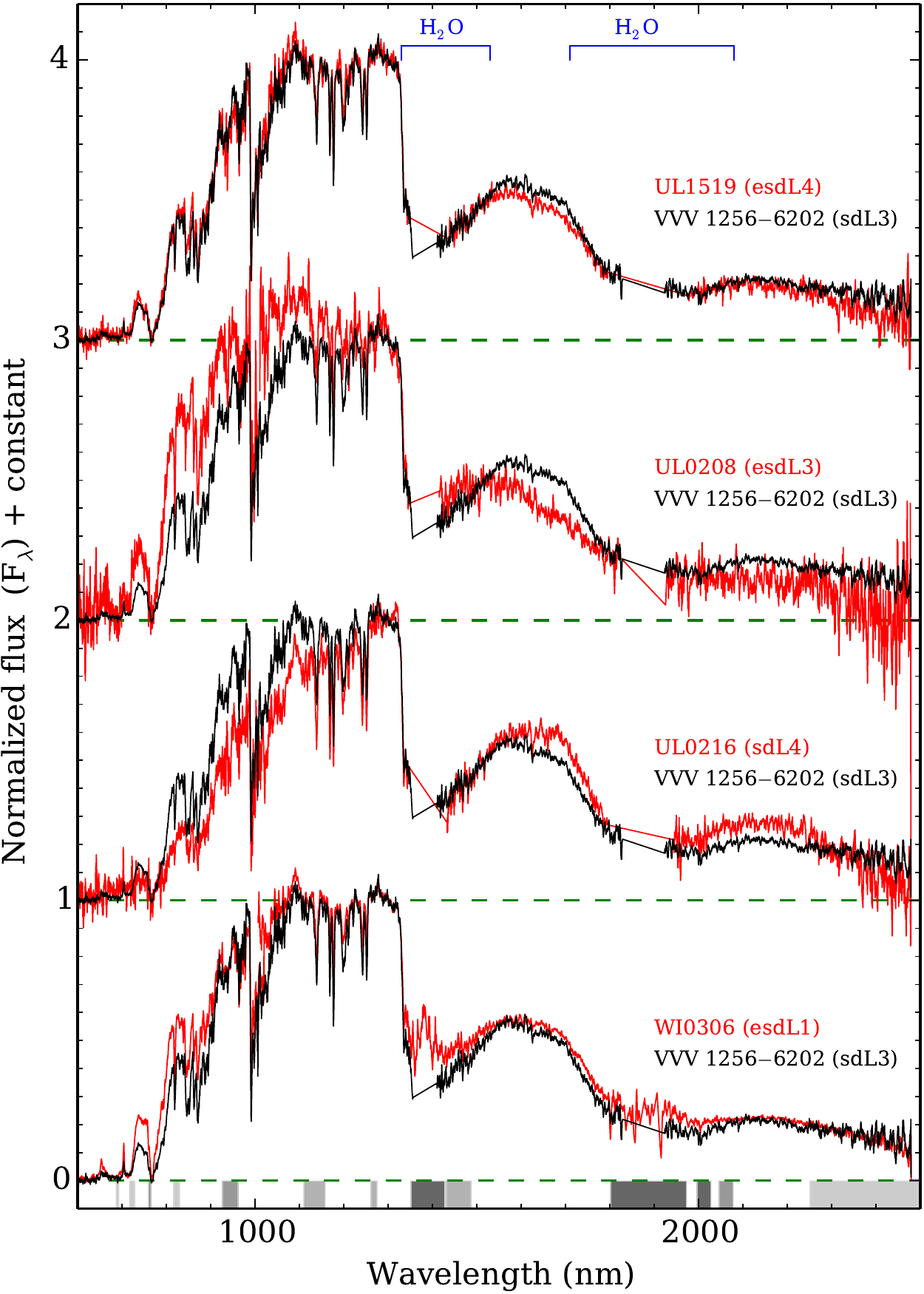}
\caption[]{Optical to NIR spectrum of VVV 1256$-$6202 compared to that of UL1519 \citep{zha17a}, UL0208 \citep{zha18a}, UL0216, WI0306, observed with X-shooter. Spectra are normalized at 1300 nm.} 
\label{fnirj}
\end{center}
\end{figure}

\begin{figure}
\begin{center}
   \includegraphics[width=\columnwidth]{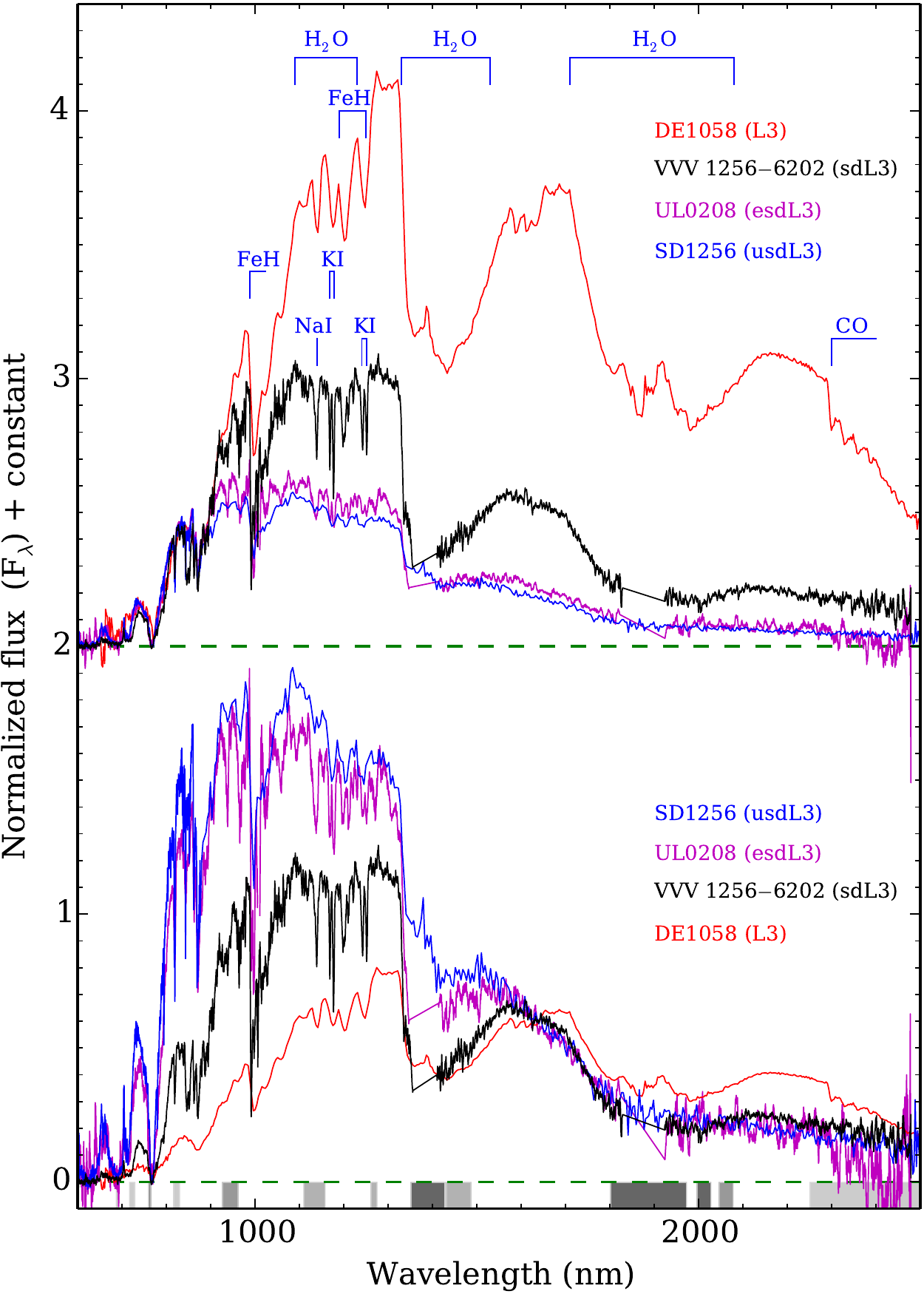}
\caption[]{Optical to NIR spectra of L3 \citep[DE1058;][]{burg10}, sdL3 (VVV 1256$-$6202), esdL3 (UL0208), and usdL3 \citep[SD1256;][]{bur09}. Spectra are normalized at 840 nm (top) and 1620 nm (bottom). } 
\label{fnirh}
\end{center}
\end{figure}

The signal-to-noise ratio (SNR) of the original spectrum is $\sim$ 12 at 740 nm, 24 at 830 nm, 28 at 920 nm, 23 at 1090 and 1300 nm, 20 at 1600 nm, and 27 at 2150 nm. The X-shooter spectrum displayed in Figs \ref{fvis}--\ref{fmod} was smoothed by 101 and 51 pixels in the VIS and NIR arms which increased the SNR by $\sim$10 and seven times, respectively, and reduced the resolving power to about 600. 

We also obtained a new optical to NIR spectrum of a previously known L subdwarfs, WISEA J030601.66$-$033059.0 \citep[WI0306;][]{kir14,luhm14}, with the X-shooter on 2015 September 19 under a seeing of 1.30 and an airmass of 1.14. The total exposure time was divided in two single integrations of 140 s for the VIS arm and 150 s for the NIR arm. A B8 II star (HIP 26623) observed right after WI0306 with an airmass of 1.07 was used for telluric correction.

\begin{table}
 \centering
  \caption[]{Properties of VVV J12564163$-$6202039.}
\label{prop}
  \begin{tabular}{l c c c}
\hline
Parameter & Value  \\	
\hline 
 {\sl Gaia} DR2 ID & 5863122429178232704 \\
 $\alpha$  & $12^{\rm h}56^{\rm m}41\fs06$ \\
 $\delta$  &  $-62\degr02\arcmin03\farcs9$ \\
 $l$ & 303\fdg5470023 \\
 $b$ & 0\fdg8310617 \\
Epoch & 2015.5 \\
Spectral type & sdL3 \\
$G$ &  20.75 \\
$\pi$ (mas) & 14.94$\pm$1.68 \\
Distance (pc) &  66.94$^{+8.49}_{-6.77}$ \\
$\mu_{\rm RA}$ (mas yr$^{-1}$) &   $-$1129.81$\pm$3.25   \\
$\mu_{\rm Dec}$ (mas yr$^{-1}$) &  21.29$\pm$2.56  \\
$V_{tan}$ (km s$^{-1}$) &  359$^{+45}_{-36}$ \\
RV (km s$^{-1}$) &  $-$47.6$\pm$7.7  \\ 
$U$ (km s$^{-1}$) & $-325^{+38}_{-31}$ \\
$V$ (km s$^{-1}$) & $-158^{+26}_{-21}$  \\
$W$ (km s$^{-1}$) & $13.4^{+1.8}_{-1.4}$ \\
$T_{\rm eff}$ (K) & 2220$\pm$100 \\
${\rm [Fe/H]}$ & $-$0.9$\pm$0.2 \\
%log $g$ & 5.5$\pm$0.25 \\
\hline
\end{tabular}
\end{table}

\section{Characterization}
\label{scha}

\subsection{Spectral classification}
\label{ssc}
L subdwarfs have the most diverse spectral features as they have cloud formation in their atmospheres and a wide range of metallicity that is shaping their spectra. L subdwarfs are classified into usdL, esdL, and sdL subclasses corresponding to metallicity ranges of [Fe/H] $\la -1.7$, $-1.7 \la$ [Fe/H] $\la -1.0$, and $-1.0 \la$ [Fe/H] $\la -0.3$, respectively  \citep{zha17a}. Figs \ref{fvis}--\ref{fnirh} show optical and NIR spectrum of VVV 1256$-$6202 in comparison to that of an L3 dwarf standard and other known L subdwarfs. 

Optical spectra of L dwarfs are used as reference to assign subtypes of L subdwarfs \citep{bur03,burg07,kirk10}. The L3 type DENIS-P J1058.7-1548 \citep[DE1058;][]{delf97} has the closest optical profile at 730--760 and 860--925 nm wavelengths to VVV 1256$-$6202 among L dwarf standards (Fig. \ref{fvis}). The weak VO absorption band at 800 nm indicates that VVV 1256$-$6202 is an early-type sdL subdwarf \citep[table 3;][]{zha17a}. Therefore, we classified VVV 1256$-$6202 as an sdL3 subdwarf. 

We compared the new X-shooter spectrum of VVV 1256$-$6202 to its FIRE spectrum from \citet{smit18}. We found that its FIRE spectrum is consistent with the X-shooter spectrum at 1700--2400 nm wavelength. However, the flux calibration of its FIRE spectrum at 800--1400 nm wavelength was not correct, and that misled its initial spectral classification.

%which is consistent with the result inferred from the comparison to UL1519 and SD1333. 

The TiO absorption band at around 850 nm in spectra of early- and mid-type L subdwarfs is a good indicator of metallicity but its strength is not a monotonous function of metallicity \citep[fig. 10;][]{zha17a}. Fig. \ref{ftio} shows that the 850 nm TiO absorption band is strengthening from the L3 type DE1058 to the sdL4 type ULAS J144151.55+043738.5 \citep[UL1441;][]{zha18b}, and reached to its maximum in the sdL3 type VVV 1256$-$6202. Then it is weakening across the esdL4 type ULAS J151913.03$-$000030.0 \citep[UL1519;][]{zha17a}, the esdL3 type ULAS J020858.62+020657.0 \citep[UL0208;][]{zha18a}, and the usdL3 type SDSS J125637.13-022452.4 \citep[SD1256;][]{siv09}. 

VVV 1256$-$6202 has slightly stronger 850 nm TiO absorption than the sdL1 type SDSS J133348.24+273508.8 \citep[SD1333;][]{zha17a} which is also close to the boundary between sdL and esdL subclasses and has halo kinematics. However, they possibly have a similar metallicity, as the 790 nm VO absorption band in VVV 1256$-$6202 is as weak as that in SD1333. The slight difference on the strength of their 850 nm TiO absorption might be related to their $T_{\rm eff}$ differences, as SD1333 is two subtypes earlier. A NIR spectrum of SD1333 is required to place a tighter constraint on its metallicity.  

The metallicity and $T_{\rm eff}$ degeneracy is common in the spectral classification of L subdwarfs. Both optical and NIR spectra are usually required to break the metallicity and $T_{\rm eff}$ degeneracy and assign the subclass and subtype of L subdwarfs.

Fig. \ref{fnirj} shows differences between optical to NIR spectra of VVV 1256$-$6202 and four known L subdwarfs with closest NIR spectra. UL1519 has the closest optical and NIR spectral profile to VVV 1256$-$6202 (Fig. \ref{fnirh}). However, UL1519 (esdL4) should have a lower metallicity than VVV 1256$-$6202, because UL1519 has stronger $H$ band flux suppression caused by enhanced collision-induced H$_2$ absorption (CIA H$_2$) at lower metallicity and weaker 850 nm TiO absorption band than VVV 1256$-$6202. Spectra of L subdwarfs are redder at higher metallicity or lower $T_{\rm eff}$, and bluer at lower metallicity or higher $T_{\rm eff}$. Therefore, VVV 1256$-$6202 should have a higher $T_{\rm eff}$ (and earlier spectral type) than UL1519. This is also indicated by its relatively higher flux at 805-835 nm wavelength. VVV 1256$-$6202 and UL1519 coincidentally have similar optical to NIR spectra, as the influences from their $T_{\rm eff}$ and metallicity differences pretty much cancelled each other out in their spectra. 

%We argued that the spectral difference of VVV 1256$-$6202 from UL1519 is due to a slightly higher metallicity and higher $T_{\rm eff}$ than UL1519. 
VVV 1256$-$6202 has a relatively redder spectrum than UL0208 that indicates it has a higher metallicity than UL0208. VVV 1256$-$6202 has flatter $K$ band spectrum than UL0216 that indicates it has a lower metallicity than UL0216. Its bluer overall spectrum than UL0216 suggests that VVV 1256$-$6202 has a higher $T_{\rm eff}$ than UL0216. VVV 1256$-$6202 has much redder optical spectrum than WI0306 which is classified as esdL1 \citep{zha17a}, as VVV 1256$-$6202 is two subtypes later and also has slightly higher metallicity (stronger VO absorption at 800 nm) than WI0306. However, they have similar $J, H, K$ band spectral profile, but VVV 1256$-$6202 clearly has deeper water absorption bands. 

The top panel of Fig. \ref{fnirh} shows the optical to NIR spectra of DE1058 (L3), VVV 1256$-$6202 (sdL3), UL0208 (esdL3), and SD1256 (usdL3) normalized in the optical. VVV 1256$-$6202 has significant suppressed NIR flux (caused by enhanced CIA H$_2$) than the L3 type DE1058, but not as much as the esdL3 type UL0208 and the usdL3 type SD1256. 
L subdwarfs show small variations on their surface gravity. Their spectral type is mainly composed of two parameters. The subtype is an indicator of $T_{\rm eff}$ (and clouds). The subclass is an indicator of metallicity \citep{kirk05}.  
The metallicity range of L subdwarfs in each subclasses (sdL, esdL, usdL) is consistent across subtypes  under the L subdwarf classification scheme of \citet{zha17a}. However, the $T_{\rm eff}$, luminosity, and absolute magnitude of each individual subtype are not consistent across different subclasses. Therefore, these spectra in the top panel of Fig. \ref{fnirh} do not show the true relative fluxes between the dL3, sdL3, esdL3, and usdL3 subclasses at the same distance. L3 subdwarfs are actually more massive and hotter than L3 dwarfs \citep{zha18a}. Fig. 22 in \citet{zha18b} shows that L3 dwarfs and subdwarfs have similar $H$ band absolute magnitude. Therefore, a set of L3, sdL3, esdL3, and usdL3 spectra normalized at $H$ band, as shown in the bottom panel of Fig. \ref{fnirh}, would indicate their relative fluxes at the same distance. 

\subsection{Atmospheric properties}
%\label{spro}
%The BT-Settl/BT-Dusty models \citep{alla11,alla14} are the very successful in reproducing spectral features ultracool subdwarfs, which have relatively simpler atmospheres (less molecules) at subsolar metallicity \citep{zha17a}. However,  

L subdwarfs have subsolar metallicity and relatively simpler atmospheres (less molecules) compared to L dwarfs. Observed spectra of L subdwarfs are relatively well reproduced by the latest BT-Settl/BT-Dusty models \citep{alla11,alla14,zha17a}. However, there are some difficulties in the determination of atmospheric parameters of L subdwarfs by model spectral fittings. 
First, model atmospheres are not tested at subsolar metallicity due to the lack of wide L subdwarf companions to stars with known metallicity. Secondly, some spectral features of L subdwarfs are not well reproduced by model atmospheres. For example, the 850 nm TiO absorption band in halo L subdwarfs is underestimated by models \citep{zha18a}. The $H$ band spectra of L subdwarfs with either [Fe/H] $\ga -$1 \citep{zha17a} or [Fe/H] $\la -$2 \citep{zha17b} are not well reproduced by models. Thirdly, both temperature and metallicity have large influence on spectral profile of L subdwarfs. VVV 1256$-$6202 has very similar optical to NIR spectrum as UL1519, but they have different atmospheric properties according to some detailed metallicity sensitive spectral features (Section \ref{ssc}). 

%Nevertheless, there are some well known metallicity sensitive spectral features across the optical to NIR wavelength that can be used to break the metallicity--temperature degeneracy \citep{zha17a}. 

To avoid the metallicity--temperature degeneracy problem in automatic spectral fitting of L subdwarfs, we manually fitted the spectrum of VVV 1256$-$6202 to a set of BT-Dusty model spectra primarily by these wavelength ranges that are sensitive to metallicity and/or temperature, e.g. TiO and VO absorption bands at 700-820 nm and the $J$ band flux around 1300 nm. We also ignored wavelengths that are not well reproduced by models. 
The model that we used has intervals of 100 K for $T_{\rm eff}$ and 0.5 dex for [Fe/H]. We used a fixed gravity of log $g$ = 5.5, as L subdwarfs all have very similar gravity around 5.5 dex. We applied linear interpolation between model spectra to improve fitting. The interpolated intervals are 20 K for $T_{\rm eff}$ and 0.1 dex for [Fe/H].

\begin{figure}
\begin{center}
   \includegraphics[width=\columnwidth]{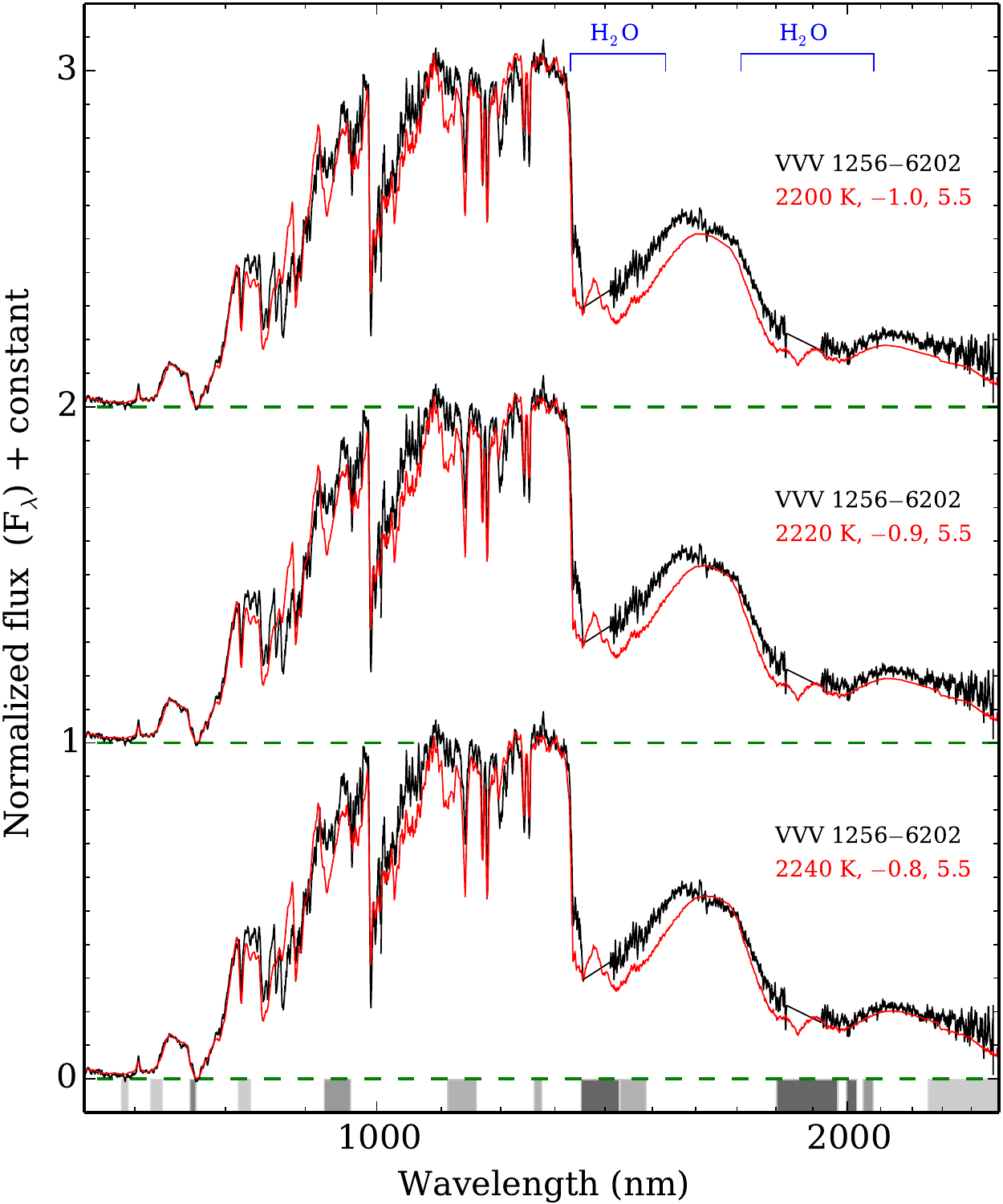}
\caption[]{Optical to NIR spectrum of VVV 1256$-$6202 and its best-fitting BT-Dusty model spectra. Spectra are normalized at 1300 nm.} 
\label{fmod}
\end{center}
\end{figure}

Fig. \ref{fmod} shows the X-shooter spectrum of VVV 1256$-$6202 compared to its best-fitting BT-Dusty model spectra. These models well reproduced observed spectrum at 650--1350 nm wavelength but overestimated the strength of the water absorption band at around 1500 nm. The model with $T_{\rm eff}$ = 2200 K and [Fe/H] = $-1.0$ generally reproduced the spectral shape at 1600--2500 nm but slightly underestimated the flux level at $H$ and $K$ bands. The model with $T_{\rm eff}$ = 2240 K and [Fe/H] = $-0.8$ reproduced the flux level at $K$ band but does not have an accurate $H$ band spectral shape. It has stronger water absorption at around 1500 and 1800 nm than the observed spectrum. The model with $T_{\rm eff}$ = 2220 K and [Fe/H] = $-0.9$ has an intermediate $H$ and $K$ band profile between the other two models. The uncertainties of the model fitting are slightly larger than the interpolated intervals. BT-Dusty model with subsolar metallicity has not been tested with benchmarks, e.g. BD + star binaries with known metallicity and age \citep{maro17}, thus has potentially systematic uncertainty on predicted atmospheric parameters. The total uncertainties of $T_{\rm eff}$ and [Fe/H] of VVV 1256$-$6202 should be $\leq$ 100 K and $\leq$ 0.2 dex, respectively. The metallicity of VVV 1256$-$6202 derived from the models agrees with its sdL spectral classification.

The $T_{\rm eff}$ and [Fe/H] derived from model spectral fitting could be used to assess the substellar status of old ultracool subdwarfs. VVV 1256$-$6202 is around the stellar/substellar boundary with $T_{\rm eff} = 2220\pm100$ K and [Fe/H] = $-0.9\pm0.2$ according to fig. 9 of  \citet{zha17b}. 
The stellar/substellar boundary across different metallicity in fig. 9 of  \citet{zha17b} is deduced from the hydrogen burning minimum mass of 0.083 M$_{\sun}$ at [M/H] = $-$1 (i.e. [Fe/H] = $-$1.3; \citealt{chab97}) and the mass--$T_{\rm eff}$ correlations \citep{bara97}. Note that the mass--$T_{\rm eff}$ correlations with [Fe/H] = $-1.3$ and $-0.7$ intersect at around 0.084 M$_{\sun}$ where unsteady hydrogen fusion occurs \citep[fig. 8;][]{zha17b}. 
The substellar boundary near [Fe/H] = $-$1 might be slightly underestimated as those 10 Gyr iso-mass contours nearby have turned to the low $T_{\rm eff}$ side with decreasing [Fe/H], that is not the case for stars. VVV 1256$-$6202 is just below the empirical stellar--substellar boundary in $i-J$ versus $J-K$ plot \citep[fig. 17;][]{zha18b}. Therefore, VVV 1256$-$6202 is likely a T-BD.

%From model spectral fitting of VVV 1256$-$6202 we can see that the BT-Dusty model still has space to improve. Some detailed spectral features at 820--1000 nm wavelength are not correctly reproduced by the latest models \citep[e.g. flux at 830 nm, 850 nm TiO absorption;][]{zha18a}. The $H$ band spectral profile is also not correctly reproduced for L subdwarfs with [Fe/H] $\ga -1.0$ (e.g. VVV 1256$-$6202) or [Fe/H] $\la - 2.0$ \citep[e.g. SDSS J010448.46+153501.8;][]{zha17b}.   

\begin{figure*}
\begin{center}
   \includegraphics[width=\textwidth]{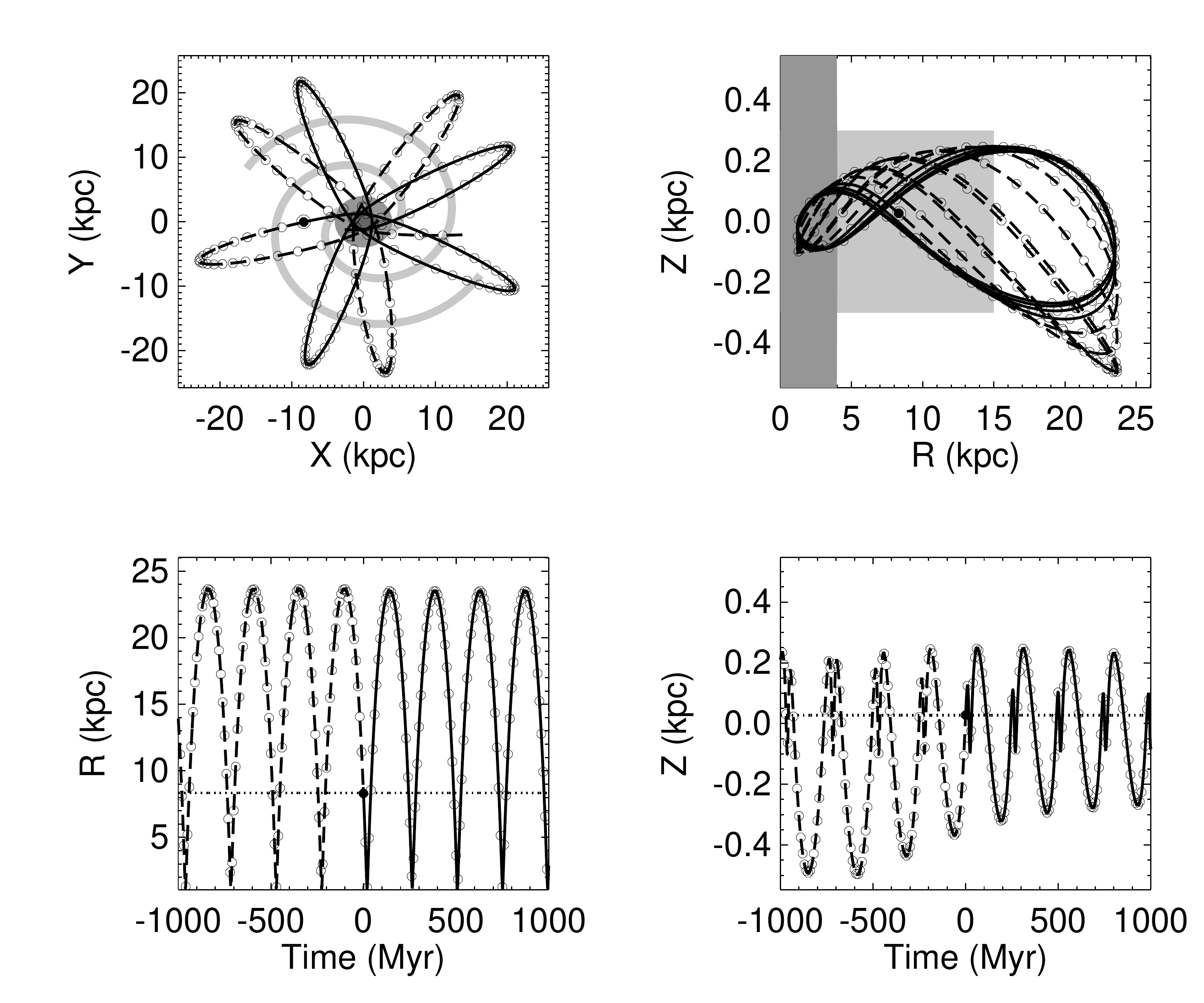}
\caption[]{Integrated prograde Galactic orbit of VVV 1256$-$6202 over a period from 1 Gyr before (dashed line) to 1 Gyr after (solid line) current epoch (black dot). The top left panel shows the projected orbit in the plane of the Galaxy [$X, Y$]. The top right panel shows the projected orbit perpendicular to the Galactic disc and in the radial direction [$R, Z$]. The darker and lighter grey shaded areas in top panels indicate the Galactic bulge and spiral arms. The bottom panels show the time evolution of the orbit in galactocentric cylindrical coordinates [$R, Z$]. }
\label{fgo}
\end{center}
\end{figure*}

\subsection{Galactic orbit}
%\label{skin}
To measure the RV of L subdwarfs based on their X-shooter spectra, we also observed three L dwarf RV standards with X-shooter \citep{zha18a,zha18b}. Since L subdwarfs and dwarfs have a lot absorption lines in comment, thus we can use the {\scriptsize IRAF} package {\scriptsize FXCOR} task to measure the RV difference between our L subdwarfs and L dwarf RV standards. Barycentric velocities are corrected for both L subdwarfs and RV standards, before we use RV standards as reference to calibrate the RV of L subdwarfs. DENIS-P J144137.3$-$094559 \citep[DE1441;][]{mart99} is one of these three L dwarf RV standards that we observed with X-shooter \citep{zha18a}. It is an L0.5 dwarf with RV of $-27.9\pm1.2$ km s$^{-1}$ \citep{bail04}. It is the one with the closest spectral profile to VVV 1256$-$6202, thus is used as the reference to measure the RV of VVV 1256$-$6202. The final RV of VVV 1256$-$6202 is $-$47.6$\pm$7.7 km s$^{-1}$. 

We calculated the $U, V, W$ Galactic space velocities with the RV measured from the X-shooter and {\sl Gaia} astrometry (Table \ref{prop}). We also calculated the kinematic membership probability of VVV 1256$-$6202 based on its $U, V, W$ space velocities and population fractions of thin-disc (0.93), thick-disc (0.07), and halo (0.006) stars in the solar neighbourhood \citep{redd06}. The kinematic probability for VVV 1256$-$6202 to be a halo or thick disc member is 99.82 and 0.18 per cent, respectively. The halo membership of  VVV 1256$-$6202 is robust considering that its metallicity ([Fe/H] $\approx -0.9$) is in the range of halo population. 

The majority of sdL subdwarfs are kinematically associated with the Galactic thick disc. However, there are three halo members among the 39 known subdwarfs of the sdL subclass. They are VVV 1256$-$6202, SD1333, and ULAS J021258.08+064115.9 \citep[UL0212;][]{zha18b} which also have the strongest metal-poor features in sdL subclass. Meanwhile, the esdL and usdL subclasses are all kinematically associated with the Galactic halo. Note that metallicity ranges of stellar populations of the thick disc and halo have a small overlap at around [Fe/H] = $-$1.0 \citep[fig. 2;][]{spag10}. 

VVV 1256$-$6202 has unusual Galactic kinematics as it is currently located in the Galactic plane ( $b$ = 0\fdg831), and has very low velocity perpendicular to the Galactic plane ($W = 13.4$ km s$^{-1}$) and very high total space velocity (362 km s$^{-1}$). This indicates that it is a halo member but orbiting in the Galactic plane. We calculated its probable Galactic orbit to have a clear view of its motion in the Milky Way following a method described in section 3.4 of \citet{bur09}. The calculation is based on its current (epoch 2015.5) kinematics ($U, V, W$ velocity), distance, and coordinates from {\sl Gaia} DR2 (Table \ref{prop}). 

Fig. \ref{fgo} shows the integrated Galactic orbit of VVV 1256$-$6202. It is now moving towards the Galactic centre. It has extremely eccentric (e = 0.9) prograde orbit that takes it from as close as 1.2 kpc from the Galactic centre to 24 kpc out. Its orbit is unusually flat and can be fitted in a disc-like space with a thickness of $\la$ 1 kpc and a diameter of 48 kpc. However, we still cannot completely trust its past and future orbit calculated based on its current kinematics and location. Because once every $\sim$0.25 Gyr, VVV 1256$-$6202 passes through the Galactic bulge and spiral arms which are the densest areas of stars in the Milky Way, it could have gravitational interactions with field stars that are passing by. Fortunately, it has very high relative velocity from field stars (e.g. $\sim$360 km s$^{-1}$ compared to $\sim$20 km s$^{-1}$ for field stars in the solar neighbourhood) that reduces the duration of possible perturbations by field stars of the order of a magnitude. Moreover, the closer VVV 1256$-$6202 is to the Galactic centre the faster it moves. VVV 1256$-$6202 spends about two-thirds of its time further out of the stellar disc from the Galactic centre ($>$15 kpc). 

VVV 1256$-$6202 is most likely a halo member according to kinematics statistics. However, we cannot rule out the possibility that it was ejected from the Galactic bulge which also contains metal-poor stars, because the orbit of VVV 1256$-$6202 is almost in the radial direction of the Milky Way and passes through the bulge. We note that the current total space velocity of VVV 1256$-$6202 is around half of that of known hypervelocity stars which are also much more massive.

\section{Conclusion}
\label{scon}
We presented a new X-shooter spectrum of VVV 1256$-$6202 discovered with the VVV survey and re-classified it as an sdL3 subdwarf. It has the strongest TiO absorption band at around 850 nm among known L subdwarfs. The strength of the 850 nm TiO absorption band is not a monotonous function of metallicity, it is strengthening from dL to sdL subclass and then weakening from sdL to esdL and usdL subclass. 

The BT-Dusty model spectral fitting shows that VVV 1256$-$6202 has $T_{\rm eff} = 2220\pm100$ K and [Fe/H] $= -0.9\pm0.2$ that place it at around the stellar/substellar boundary in the $T_{\rm eff}$ versus [Fe/H] space. Its $i-J$ versus $J-K$ colours suggest that it is likely a T-BD just below the stellar/substellar boundary. 

We calculated the $U, V, W$ space velocity of VVV 1256$-$6202 based on its RV measured from its X-shooter spectrum and {\sl Gaia} astrometry. The kinematics shows that VVV 1256$-$6202 is a member of the Galactic halo. There are only three known L subdwarfs in the sdL subclass who have halo kinematics including VVV 1256$-$6202, SD1333, and UL0212. 

VVV 1256$-$6202 has unusually flat Galactic orbit. The thickness of its prograde orbit is similar to that of stellar disc ($\sim$ 600 pc). Its orbit is extremely eccentric (e = 0.9) and takes it from as close as 1.2 kpc from the Galactic centre to 24 kpc out. It spends one-third of its time in the stellar disc crosses the Galactic bulge and spiral arms and two-thirds of its time further out from the Galactic centre ($>$ 15 kpc). VVV 1256$-$6202 is most likely a halo member by its kinematics but also could be an ejecta of the Galactic bulge.

\section*{Acknowledgements}
Based on observations collected at the European Organisation for Astronomical Research in the Southern Hemisphere under ESO programmes 0101.C-0626, 096.C-0130, 095.C-0878 and 094.C-0202. This work presents results from the European Space Agency (ESA) space mission {\sl Gaia}. {\sl Gaia} data is being processed by the {\sl Gaia} Data Processing and Analysis Consortium (DPAC). Funding for the DPAC is provided by national institutions, in particular the institutions participating in the {\sl Gaia} MultiLateral Agreement (MLA). The {\sl Gaia} mission website is \url{https://www.cosmos.esa.int/gaia}. The {\sl Gaia} archive website is \url{https://archives.esac.esa.int/gaia}. This research has benefited from the SpeX Prism Spectral Libraries, maintained by Adam Burgasser at \url{http://www.browndwarfs.org/spexprism}. Synthetic spectra used in this paper are calculated by Derek Homeier based on BT-Dusty model atmospheres created by France Allard. ZHZ is supported by the PSL Fellowship.

\bibliographystyle{mnras}
\bibliography{vvv1256} % if your bibtex file is called 

\label{lastpage}
\end{document}